\documentclass[a4paper,man,natbib,floatsintext]{apa6}
\usepackage{natbib}

\usepackage{csquotes}
\usepackage{xcolor}
\usepackage{tabularx}
\usepackage{makecell}
\usepackage{graphicx}
\usepackage{subfig}
\usepackage{verbatim}

\usepackage{caption}
\usepackage[font=footnotesize]{caption}
\usepackage{amsmath}
\usepackage{lipsum} 
\usepackage{hyperref}
\usepackage{multirow}

\usepackage{adjustbox}
\usepackage{longtable}
\usepackage[pagewise]{lineno}
\begin{document}
%

\begin{titlepage}

\begin{center}

\large Towards Context-Aware Modeling of Situation Awareness in Conditionally Automated Driving
\\ 

\normalsize

\vspace{25pt}
Lilit Avetisyan\\
Industrial and Manufacturing Systems Engineering, University of Michigan-Dearborn\\
\vspace{15pt}
X. Jessie Yang\\
Industrial and Operations Engineering, University of Michigan-Ann Arbor\\
\vspace{15pt}
Feng Zhou\\
Industrial and Manufacturing Systems Engineering, University of Michigan-Dearborn\\
\vspace{15pt}
\end{center}
\begin{flushleft}
\textbf{Manuscript type:} \textit{Research Article}\\
\textbf{Running head:} \textit{Context-Aware Situation Awareness in Automated Driving}\\
\textbf{Word count:} 

\textbf{Corresponding author:} 
Feng Zhou, 4901 Evergreen Road, Dearborn, MI 48128, Email: fezhou@umich.edu


\end{flushleft}

\end{titlepage}
\shorttitle{}



\section{ABSTRACT}
Maintaining adequate situation awareness (SA) is crucial for the safe operation of conditionally automated vehicles (AVs), which requires drivers to regain control during takeover (TOR) events. 
This study developed a predictive model for real-time assessment of driver SA using multimodal data (e.g., galvanic skin response, heart rate and eye tracking data, and driver characteristics) collected in a simulated driving environment. Sixty-seven participants experienced automated driving scenarios with TORs, with conditions varying in risk perception and the presence of automation errors. A LightGBM (Light Gradient Boosting Machine) model trained on the top 12 predictors identified by SHAP (SHapley Additive exPlanations) achieved promising performance with $RMSE=0.89, MAE=0.71$, and $Corr=0.78$. 
These findings have implications towards context-aware modeling of SA in conditionally automated driving, paving the way for safer and more seamless driver-AV interactions.


\textbf{Keywords:} Situation awareness, real-time prediction, physiological measures, eye tracking, SHAP, autonomous vehicles.


\newpage
\section{Introduction}

The rapid advancement of automated vehicle (AV) technologies holds the promise of transforming transportation. Yet, as vehicles progress through the levels of automation set by the Society of Automotive Engineers (SAE), they reach an intermediate phase known as conditionally automated driving -- SAE Level 3 \citep{sae2021}. In this phase, drivers must be ready to retake control in critical situations after receiving a takeover request (TOR) \citep{zhou2019takeover, ayoub2019manual, zhou2021using}. Research consistently demonstrates that delayed and ineffective driver response during takeovers, especially when distracted by non-driving related tasks (NDRTs), substantially harm driving safety \citep{du2019examining, du2020examining, du2020psychophysiological}. 

In conditionally automated vehicles, the maintenance of driver Situation Awareness (SA) -- the accurate perception, comprehension, and projection of environmental elements \citep{endsley1995} -- is paramount. The criticality of SA in ensuring timely and effective driver response during takeovers is well documented \citep{salmon2006situation}. However, driver engagement in NDRTs can lead to degradation of SA, leading to failures in takeovers \citep{korber2018introduction, endsley1995out}.

Risk perception, closely interwoven with SA, influences a driver's subjective assessment of potential threats, directly affecting trust in automation, vigilance levels, and takeover decisions \citep{hulse2018perceptions, SEPPELT201966}. However, the existing literature focuses predominantly on the behavioral ramifications of risk perception and SA during manual driving, with less emphasis on their interplay within the scope of conditional automation where driver disengagement factors are distinct and heightened \citep{pop2015individual, khastgir2017risk}.

Given the transient dynamics of SA in the driving context, traditional SA measurement methods, while robust, prove inadequate for capturing the continuous evolution of SA, due to their intrusive nature and reliance on self-reports, such as freeze-probe techniques (e.g., SAGAT) or observer rating scales (e.g., SART) \citep{de2019situation, durso1998situation}.  This knowledge gap highlights the need for developing unobtrusive, real-time methodologies that utilize objective indicators -- such as physiological signals and eye-tracking data -- to assess SA.

Notably, advancements in machine learning offer promising avenues for the prediction and real-time monitoring of SA in automated driving, with studies indicating potential correlations between driver's SA levels and various physiological and behavioral markers \citep{zhou2021using, du2020predicting}. Yet, research lacks a comprehensive analytical framework capable of integrating multimodal datasets to reliably predict SA, whilst considering individual differences and fluctuating driving conditions \citep{perello2021driver, smith2023physiological}.  

To bridge these gaps, the present study leverages a multimodal dataset, encompassing physiological responses and eye-tracking metrics, augmented by individual driver characteristics, to develop and validate a predictive model of SA tailored to automated driving scenarios. Specifically, a driving simulator experiment was conducted with 67 participants who experienced TOR events under varying risk perception and automation reliability conditions. Physiological data, i.e, GSR (Galvanic Skin Response), HR (Heart Rate), and HRV (Heart Rate Variability) were recorded using wearable sensors. Additionally, participants' eye movements were tracked to extract metrics such as fixation numbers, fixation duration, and dispersion across areas of interest like the center, left and right sides of road scene, NDRT display and odometer. Self-reported SA ratings were collected every 30 seconds during the drives as a ground truth.

Further, this study employs Light Gradient Boosting Machine (LightGBM) \citep{ke2017lightgbm} and SHapley Additive exPlanations (SHAP) \citep{lundberg2020local, NIPS2017_7062} to not only predict SA but also unpack the contribution of each feature to the model's decisions. In doing so, it endeavors to contribute to the design of context-aware SA monitoring systems that enhance the safety and efficiency of driver-AV symbiosis.

The contributions of this paper are thus summarized as:
\begin{itemize}
    \item Development of a non-intrusive LightGBM model that leverages multimodal sensor data for real-time SA assessment in conditionally automated driving.
    \item Identification and analysis of key physiological and behavioral predictors for SA using SHAP values.
    \item Exploration of the interplay between risk perception, driver characteristics, and SA in the context of conditional automated driving.
    \item Demonstration of the practicality and effectiveness of applying machine learning for SA prediction to foster improved driver-AV interaction.
\end{itemize}

\section{Related Work}
\subsection{Importance of SA in Conditionally Automated Driving}
Situation awareness (SA), defined as ``the perception of the elements in the environment within a volume of time and space, the comprehension of their meaning, and the projection of their status in the near future'' \citep{endsley1988design}, plays a critical role in ensuring safe and efficient interactions between humans and AVs. Drivers need to maintain sufficient SA to understand the current driving context, anticipate potential hazards, and respond appropriately to changing situations, especially during the takeover transition process. However, the unique characteristics of conditionally automated driving environments present new challenges to SA, including mode confusion and automation complacency. 

Mode confusion: Drivers may struggle to understand the current operational mode (manual vs. automated) and their respective roles and responsibilities. This uncertainty can lead to inappropriate actions and delayed responses in critical situations \citep{merat2009drivers}.
Automation complacency: Overreliance on automation can lead to decreased vigilance and reduced monitoring of the driving environment, resulting in impaired SA and increased accident risk \citep{thill2014apparent}.
Delayed takeover transitions: Smooth and timely transfer of control between the driver and the automated system is crucial for maintaining SA and avoiding accidents. Difficulty in resuming control can lead to confusion and errors, especially in complex or unexpected situations \citep{zhou2022predicting}.

\subsection{The Impact of Risk Perception on SA}

Risk perception, the subjective judgment of the likelihood and severity of potential hazards, plays a crucial role in influencing SA. Drivers who perceive a higher level of risk tend to be more vigilant and attentive to their surroundings, leading to enhanced SA. Conversely, low risk perception can lead to complacency and reduced awareness of potential dangers, ultimately impairing SA and increasing the likelihood of accidents. Several factors influence risk perception in automated driving:

Automation reliability: Drivers' trust in the automated system's capabilities significantly impacts their risk perception. If the system is perceived as reliable, drivers may be less vigilant and have lower SA. Conversely, experiences with system failures or unexpected behavior can increase perceived risk and enhance SA \citep{merat2009drivers, thill2014apparent}.
System transparency: Understanding how the automated system works and its limitations is crucial for accurate risk assessment. A lack of transparency can lead to uncertainty and distrust, negatively affecting risk perception and SA \citep{schwarz2019,manchon2022does}.
Hazard levels: The presence of potential hazards in the driving environment naturally influences risk perception. Drivers are more likely to be aware of and attentive to situations with higher perceived risk, leading to increased SA \citep{niu2022driver}.
Individual differences: Drivers' age, experience, personality traits, and cultural background can influence their risk perception and subsequently, their SA. For example, older drivers may have lower risk tolerance and exhibit higher levels of SA compared to younger drivers \citep{li2019no}.

\subsection{Measuring SA: From Subjective Assessments to Objective Measures}
Accurately assessing SA in conditionally AVs is essential for understanding driver behavior and developing effective safety interventions. Traditional SA measurement methods primarily rely on subjective assessments.
Situation Awareness Global Assessment Technique (SAGAT) is widely used method that involves freezing the driving scenario and asking participants a series of questions about the current situation to assess their level of awareness \citep{endsley1995}.
Situation Awareness Rating Technique (SART) involves trained observers rating an individual's SA based on their performance in a simulated driving task, considering factors such as scanning behavior, response times, and decision-making \citep{taylor2017sart}.
Situation Present Assessment Method (SPAM) method combines subjective self-assessments via questionnaires with objective performance measures, such as reaction times and driving errors, to provide a more comprehensive evaluation of SA \citep{durso1998situation}.
While these subjective measures offer valuable insights, they suffer from limitations such as susceptibility to biases, reliance on memory and self-perception, and inability to capture the dynamic fluctuations of SA over time \citep{salmon2006situation}. To overcome these limitations, researchers have explored the use of objective and continuous measures based on physiological and eye-tracking data:
Electroencephalography (EEG): By measuring brainwave activity, EEG can provide insights into cognitive workload, attentional focus, and mental fatigue, all of which are related to SA \citep{fernandez2019eeg, yeo2017eeg}.
Heart Rate Variability (HRV): Variations in heart rate reflect changes in the autonomic nervous system, which is influenced by cognitive and emotional states relevant to SA \citep{perello2021driver}.
Galvanic Skin Response (GSR): This measure reflects changes in skin conductance, which is associated with arousal and emotional responses that can indicate changes in SA \citep{smith2023physiological}.
Eye-tracking metrics: By monitoring eye movements, researchers can analyze fixation duration, saccade patterns, and scan paths to understand where drivers are focusing their visual attention and how efficiently they are processing information \citep{de2019situation}.
Studies have demonstrated that these objective measures can provide more accurate and reliable assessments of SA compared to subjective methods, particularly in dynamic and complex driving environments. For a comprehensive review of physiological measurements of situation awareness, please refer to \citep{zhang2023physiological}.

\subsection{Predicting SA: Towards a Multimodal and Individualized Approach}
Recent research has explored the use of machine learning and context-aware models to predict SA in real-time. These models leverage various data sources, including physiological measures, eye-tracking data, vehicle sensor data, and subjective ratings, to estimate drivers' level of awareness and predict potential lapses in attention. Some notable approaches include:
By analyzing physiological and behavioral data, deep learning algorithms can effectively predict driver SA with high accuracy. Li et al. \citeyearpar{li2022revealing} proposed a deep learning model that achieved promising results in predicting driver SA during conditionally automated driving scenarios.
Eye-tracking based models: Eye movements provide valuable insights into drivers' visual attention and information processing. Zhou et al. \citeyearpar{zhou2021using} developed a machine learning model using eye-tracking data to predict SA levels in real-time.
Functional near-infrared spectroscopy (fNIRS) allows for direct measurement of brain activity during driving tasks. Unni et al. \citeyearpar{unni2017assessing} used fNIRS to identify distinct neural signatures associated with different levels of SA.
By incorporating information about the driving environment, traffic conditions, and driver behavior, context-aware models can provide more accurate and personalized predictions of SA. Zhang et al. \citeyearpar{zheng2018effect} demonstrated the effectiveness of incorporating contextual information in predicting driver SA.
While these advancements show promise, most existing research focuses on single modalities or specific contexts. A more comprehensive approach is needed that integrates various data sources and considers individual differences to develop robust and personalized SA prediction models.

Our research aims to address the existing gaps in SA prediction by developing a multimodal and individualized approach. We will collect and analyze physiological measures, eye-tracking data, and individual characteristics to build a comprehensive model that predicts SA in conditionally automated driving scenarios. This research holds significant potential for developing real-time SA monitoring systems that can provide timely interventions and improve driver safety and performance in AVs. Our study will contribute to the growing body of knowledge on SA in automated driving and pave the way for more personalized and effective driver support systems in the future.

\section{Methodology}

\subsection{Participants} 
A total of 67 people (30 females: mean age = 28.3, SD = 11.5; and 37 males: mean age = 25.9, SD = 12.3) participated in this study. Due to malfunction of physiological sensors and the driving simulator, 23 participants were excluded, and data from the remaining 44 participants were used for further analysis. All the participants had a valid driver’s license with an average of 9.1 years of experience. Participants received \$30 in compensation for about 75 min of participation. The study was approved by the Institutional Review Board at the University of Michigan. 

\begin{figure}[tb!]
\centering
\includegraphics[width=.9\linewidth]{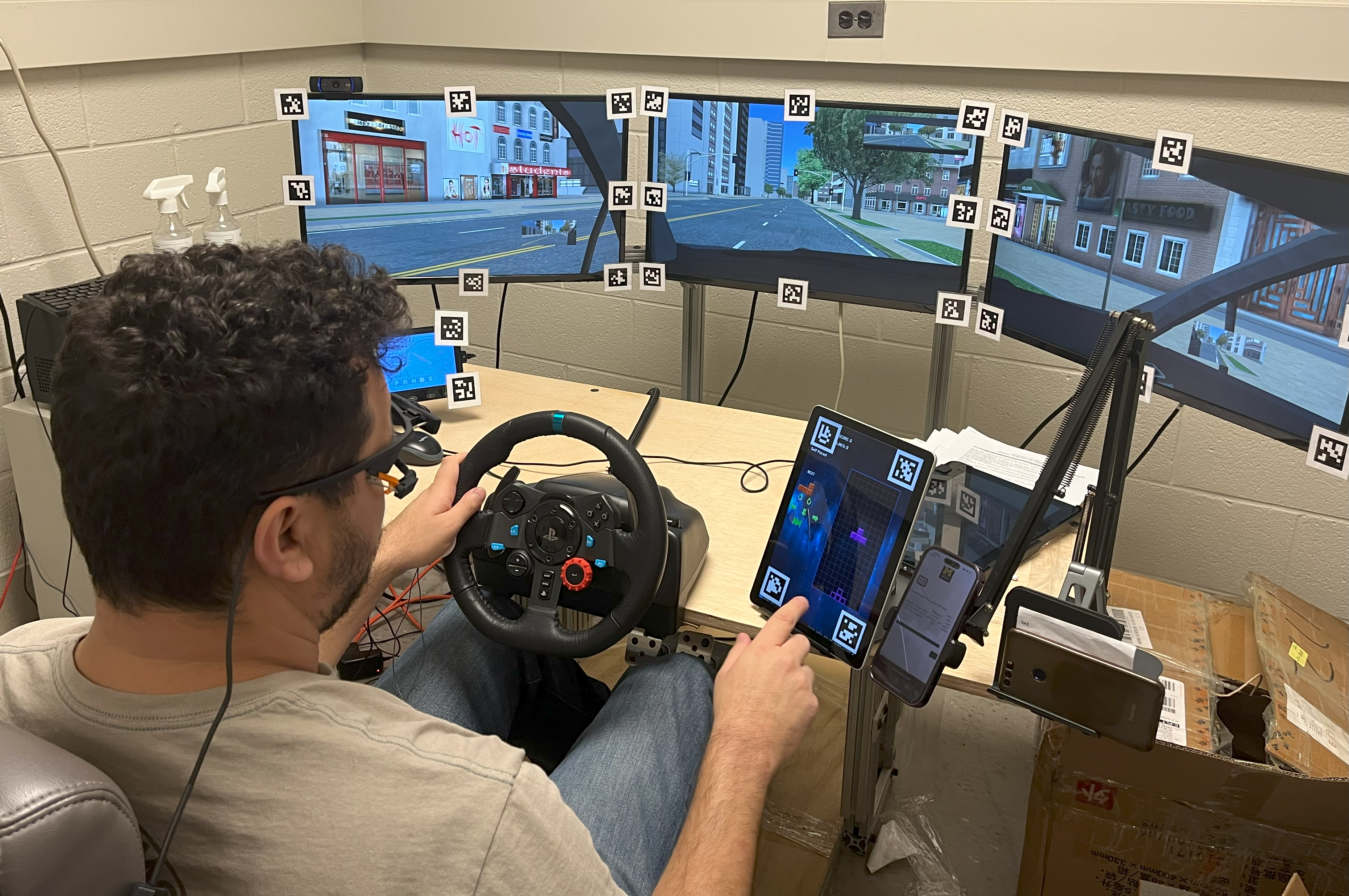}\hfill
\caption{Experiment setup.}
\label{fig:setup}
\end{figure}

\subsection{Apparatus and stimuli}
This research utilized a desktop-based driving simulator by Realtime Technologies Inc. (RTI, Michigan, USA) to gather experimental data. The simulator system included an array of three LCD monitors, a Logitech G29 driving kit, one tablet for engagement in non-driving related tasks (NDRTs), and one phone, positioned to the participant's right side, for recording SA assessments (see Fig. \ref{fig:setup}). The tablet and phone were moved to the left side for left-handed participants upon request. For the NDRT, a specially engineered Tetris game was developed using the PyGame library within the Python programming environment. The game's flow allowed participants to engage with the game tiles upon NDRT initiation, it automatically paused when TORs were triggered, enabling a seamless resumption from the previous state during the next NDRT request.


The driving simulation system was set to emulate a vehicle with conditionally automated driving (SAE Level 3) capabilities \citep{sae2021}. To engage in the automated drive mode, participants were instructed to press a red button positioned on the steering wheel. Upon the engagement of this mode, participants received a confirmation, an audio ``Automated mode engaged'' prompt, and the mode indicator on the odometer turned white. Then, the AV continued navigating a pre-defined route at a steady speed of 35 mph. While experiencing automated driving, participants were requested to start the NDRT (i.e., the Tetris game on a tablet) upon receiving the ``Please start the secondary task'' audio prompt. When a TOR -- a ``Takeover'' audio request was initiated, participants were alerted to disengage the automated mode and take manual control of the vehicle. If a participant was unable to resume vehicle control within seven seconds, 
an ``Emergency Stop'' audio alarm was activated, and the AV was triggered to stop immediately.

The self-reported SA assessment was conducted through a single-item questionnaire prompt (see Fig. \ref{fig:SA_scale}) developed on the Qualtrics platform (Provo, UT, www.qualtrics.com), and administered via a mobile phone.
The simulation also recorded physiological responses. To capture the details of visual attention, the Pupil Core eye-tracker headset with a frequency of 200 Hz from Pupil Lab (MA, USA) was used,  Concurrently, GSR and HR (via photoplethysmography or PPG) were recorded at 128 Hz using the iMotions platform with the Shimmer3 GSR+ Unit (Shimmer, MA, USA). To ensure precise synchronization of time, the time delays for each piece of equipment, including the iMotions, Pupil Core, driving simulator, and Tetris game, were recorded at the moment of the experiment's initiation.

\subsection{Experimental design}
The study employed a 2x2 mixed design experiment where the between-subjects factor was the risk condition (high-risk vs. low-risk), and the within-subjects factor was the automation errors (error vs. no error). Participants were randomly assigned to one of the risk conditions: 1) a high-risk condition, where participants were presented with negative information about AV performance through videos showcasing its malfunctions (Link: \href{https://www.youtube.com/watch?v=RC9iK1lV77E&t=4s}{high-risk video}), coupled with an environment simulation of driving in foggy weather; 2) a low-risk condition, which shared positive feedback on AV performance via videos demonstrating AV's ability to anticipate safety-critical hazards that may be difficult for humans to detect (Link: \href{https://www.youtube.com/watch?v=O4IUc0xXZqo}{low-risk video}), and featured sunny weather driving simulation. These conditions were established using video examples derived from authentic environments.

\begin{figure}[tb!]
\centering
\includegraphics[width=.7\linewidth]{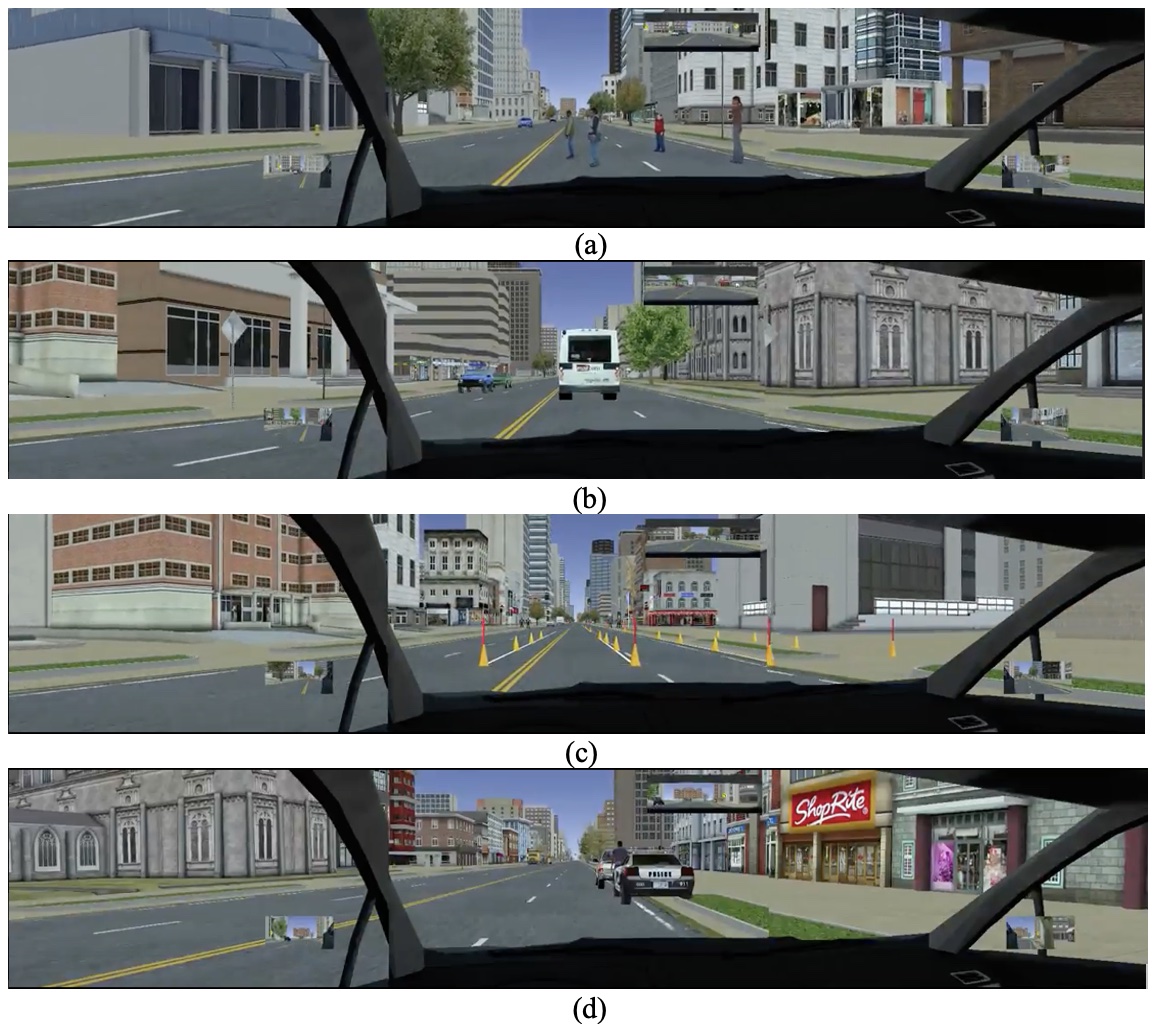}\hfill
\caption{Takeover events in urban areas (a) pedestrians crossing ahead (b) bus sudden stop ahead (c) construction zone ahead (d) police vehicle on shoulder.}
\label{fig:events2}
\end{figure}

Each participant experienced two drives with varying automation errors: 1) No Error: The AV issued four accurate TORs without technical faults, 2) Error: Participants experienced two accurate TORs (first and fourth) and two erroneous TORs (second: false alarm, third: miss). Standard road scenarios were used to trigger accurate TORs or simulate errors (e.g., pedestrians crossing, construction zones, accidents)(see Fig. \ref{fig:events2}). Previous research suggested both false alarms and misses can decrease user trust \citep{ayoub2023real}. The sequence of drives was counterbalanced across participants.


\subsection{Experimental procedure}
\begin{figure}[h!]
\centering
\includegraphics[width=.9\linewidth]{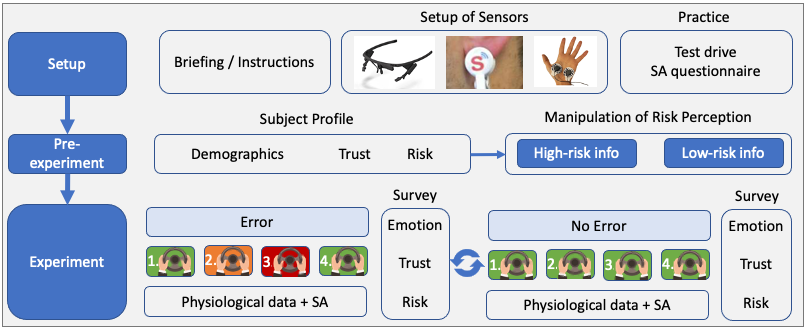}\hfill
\caption{Experiment layout.}
\label{fig:expProc}
\end{figure}

\begin{figure}[h]
\centering
\includegraphics[width=.7\linewidth]{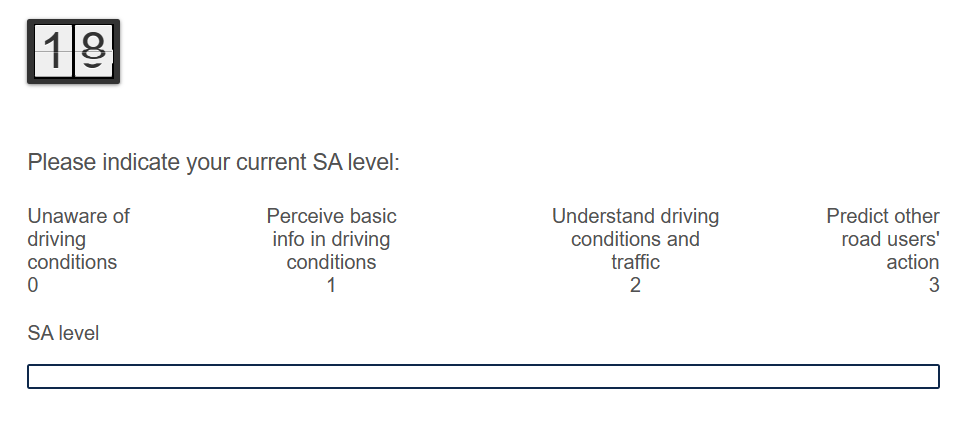}\hfill
\caption{SA level assessment prompt.}
\label{fig:SA_scale}
\end{figure}

Fig. \ref{fig:expProc} provides an overview of the experiment procedure. At the onset of the experiment, participants were briefed on the equipment and were provided with instructions regarding the tasks they would be performing. They were informed on the capabilities and limitations of Level 3 AVs, with a specific focus on the necessity for vigilance and readiness to take control whenever a TOR was issued. It was further explained that there might be instances where the AV could fail to detect an obstacle, termed as a ``miss'' condition, and situations where it might issue unnecessary TORs, referred to as ``false alarms''.
Following the briefing, participants underwent the device setup process. This involved attaching GSR electrodes to the participants' left palm and recording the baseline, fixing a PPG sensor to their left earlobe, and calibrating the eye-tracking device. Once these devices were correctly configured, participants completed an online survey to provide demographic information. Subsequently, they were presented with information specific to their assigned risk condition. 
Next, participants were guided through a training session to familiarize themselves with both the driving simulator's functionality and the experimental protocol.
Finally, they proceeded to the actual driving sessions, each lasting approximately 15 minutes. During the drive, participants were asked to self-assess and report their SA levels every 30 seconds on a 4-item scale ranging from 0 to 3, with the instruction ``Please indicate your situation awareness'' (see Fig. \ref{fig:SA_scale}). After each driving session, participants were requested to complete surveys evaluating their trust \citep{holthausen2020situational}, emotional responses \citep{jensen2020anticipated}, and perceived risk \citep{zhang2019roles} based on their recent driving experience using a 7-point Likert scale.
The total duration of the experiment was approximately 75 minutes. Note the results of emotional responses were not reported in this paper as our focus here is SA.

\section{Predictive SA Model}
\subsection{Data Preprocessing}
The study utilized a multimodal dataset comprised of physiological signals (GSR, HR, HRV), eye-tracking data, self-reported SA assessments, and demographic information. Here's the data preparation process:

GSR Processing: The GSR signal was decomposed into tonic and phasic components using the Neurokit2 package \citep{Makowski2021neurokit}. The phasic component, known for its sensitivity to rapid changes, was used for analysis.

HRV Calculation:  The HRV was calculated from collected IBI (inter-beat interval) with Root Mean Square of Successive Differences (RMSSD) method (see Eq. 1):

\begin{align*} RMSSD=&\sqrt {\sum _{i=1}^{N} \big (RR\_interval_{i}- RR\_interval_{i+1}\big)^{2}},\label{eq:rmssd} \tag{1}\end{align*}

where  $N$ is the number of heartbeats and the ``RR interval'' is the distance (in milliseconds) between two consecutive successful heartbeats

Eye-Tracking Metrics: Relevant eye-tracking metrics were extracted, including fixation numbers, fixation duration (in milliseconds), and dispersion (in degrees). Fixation numbers were identified using the I-DT algorithm \citep{salvucci2000identifying} (maximum dispersion: 1 degree, maximum duration: 200ms).

Synchronization and Alignment: Data from all sources (iMotions, Pupil Core, Qualtrics) were synchronized using timestamps across the recorded platform delays.  A 30-second sliding window was applied to calculate average GSR, HR, and eye-tracking metrics within each window. Self-reported SA ratings were linked to the physiological and eye-tracking data based on timestamps.

Integration: The final dataset integrated physiological, eye-tracking, and demographic data (age, gender, AV knowledge level), resulting in the 21 features outlined in Table \ref{tab:features}.

\subsection{LightGBM model}
This research aims to build a real-time predictive model for driver's SA using a multimodal dataset of physiological signals (i.e., GSR, HR, HRV), eye-tracking data, alongside self-reported assessments and demographic variables. After testing several ML algorithms, the LightGBM framework proved to be the most effective and was selected for this purpose. 

LightGBM is widely used in handling large datasets and high-dimensional features for regression tasks. It employs a leaf-wise growth strategy for decision trees, which can lead to better accuracy with less computation when compared to depth-wise growth strategies used by other algorithms. It is also well-suited for scenarios where speed and memory usage are critical, such as AVs, without compromising on model performance. Moreover, LightGBM supports advanced techniques like Gradient-based One-Side Sampling (GOSS) to reduce the memory usage and improve the training speed, and Exclusive Feature Bundling (EFB) to reduce the number of features and improve the efficiency of the model. Another advantage of LightGBM is that it effectively handles categorical or ordinal features directly, e.g. AV knowledge level in our dataset.  During training, LightGBM considers these  values as categorical features and looks for the best splits based on the categorical nature of the data. 

In this study, LightGBM regressor was employed to predict driver's SA. First, the model's performance was optimized using a grid search for hyperparameter tuning. Based on the tuning results, the following values set for the model parameters:  'objective': 'regression', 'metric': \{'mae', 'rmse'\}, 'learning\_rate': 0.05, 'min\_data\_in\_leaf': 20, 'num\_leaves': 50, 'early\_stopping\_rounds': 100. Next, the model was trained and validated using the dataset of 21 features and 1634 samples, and using a 10-fold cross-validation methodology. As critical indicators of model's performance, the root mean square error (RMSE) and mean absolute error (MAE) were calculated as follows:

\begin{align*} 
RMSE=&\sqrt {\frac {\sum _{i=1}^{N} \big (y_{i}-\hat {y}_{i}\big)^{2}}{N}},\label{eq:rmse} \tag{2}\\ 
MAE=&\frac {\sum _{i=1}^{N} |y_{i}-\hat {y}_{i}|}{N}, \label{eq:mae}\tag{3}\\ 
Corr.=&\frac {\sum _{i=1}^{N} (\hat {y_{i}}-\bar {\hat {y}})(y_{i} -\bar {y})} {\sqrt {\sum _{i=1}^{N} (\hat {y_{i}}-\bar {\hat {y}})^{2} \sum _{i=1}^{N}(y_{i} -\bar {y})^{2}}},\label{eq:corr}\tag{4}
\end{align*}

where N is the total number of the samples in the dataset, $y_{i}$ is the i-th value of SA samples, $\hat{y_{i}}$ is the i-th predicted SA, $\bar{y}$ is the mean value of all the SA samples, and $\bar{\hat{y}}$ is the mean value of all the predicted SA results.

\begin{table}[H]
\centering
\small
\caption{Prediction Model Features. The asterisk (*) denotes the feature's importance in the prediction model.}
\renewcommand{\arraystretch}{0.8}
\begin{tabularx}{\textwidth}{p{0.4\linewidth}p{0.08\linewidth}p{0.45\linewidth}}
\hline
Features & Unit & Description\\ 
\hline
1. age * & years  & Participant's age\\ 
2. avKnowledge * & - & Participant's knowledge level about AVs\\ 
3. gender * & -  & Participant's gender\\
4. mean\_gsr * & $\mu$S & Average galvanic skin response in phasic phase \\ 
5. mean\_HR * & bpm & Average number of heartbeats\\ 
6. mean\_HRV * & ms & Average of the variation in the time interval between heartbeats \\ 
7. number\_of\_fixations\_center * & - & Number of fixations on the center screen\\ 
8. number\_of\_fixations\_game * & - & Number of fixations on the game display \\ 
9. number\_of\_fixations\_left & - & Number of fixations on the left screen\\ 
10. number\_of\_fixations\_right & - & Number of fixations on the right screen\\ 
11. number\_of\_fixations\_odometer & - & Number of fixations on the odometer\\ 
12. mean\_dispersion\_center * & degree& Average distance between all gaze locations during a fixation on center screen\\ 
13. mean\_dispersion\_game * & degree & Average distance between all gaze locations during a fixation on game screen\\ 
14. mean\_dispersion\_left & degree & Average distance between all gaze locations during a fixation on left screen\\ 
15. mean\_dispersion\_right & degree & Average distance between all gaze locations during a fixation on right screen\\ 
16. mean\_dispersion\_odometer & degree & Average distance between all gaze locations during a fixation on odometer screen\\ 
17. mean\_duration\_center * & ms & Average duration of fixations on the center screen\\ 
18. mean\_duration\_game * & ms & Average duration of fixations on the game screen\\ 
19. mean\_duration\_left & ms & Average duration of fixations on the left screen\\ 
20. mean\_duration\_right & ms & Average duration of fixations on the right screen\\ 
21. mean\_duration\_odometer & ms & Average duration of fixations on the odometer screen\\  

\hline
\end{tabularx}
\label{tab:features}
\end{table}

\subsection{SHAP Explainer}
To reveal the contribution of each feature in the predictive model of SA, the SHAP \citep{lundberg2017unified} approach was used.  SHAP values provide a consistent and locally accurate method to attribute the effect of each feature in a prediction task, based on the foundational principles of Shapley values from cooperative game theory \citep{kuhn1953contributions}. It ensures that each feature receives an importance weight by averaging over all possible permutations of feature orderings while considering the interaction effects between features.
Moreover, SHAP model provides a detailed explanation for local/individual predictions, as well as aggregate SHAP values across multiple instances, offering a broader view of feature importance and model behavior. 
As the SA prediction model training process used a 10-fold cross-validation, the SHAP values were calculated ten times, once for each fold. The final impact of each feature was then determined by the average of these ten sets of SHAP values, providing a measure of the average contribution over the entire cross-validation process \citep{ayoub2023real, ayoub2021modeling}.

\section{Results}

\subsection{Model performance}
\begin{figure}[ht!]
\centering
\includegraphics[width=.7\linewidth]{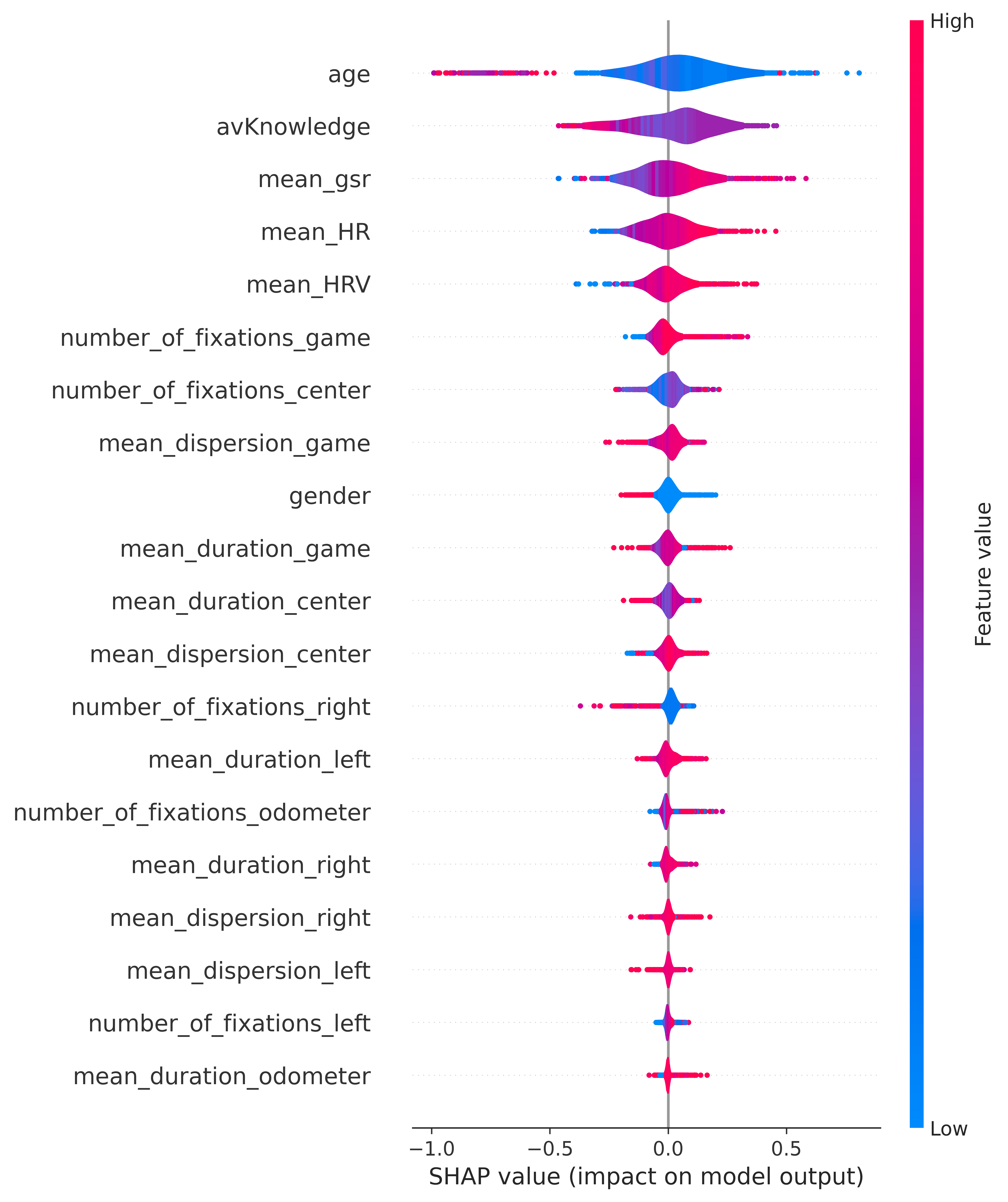}\hfill
\caption{SHAP summary plot. The x-axis shows the feature's influence on SA. The y-axis shows the importance ranking of the features.}
\label{fig:SHAP}
\end{figure}

\begin{figure}[ht!]
\centering
\includegraphics[width=.7\linewidth]{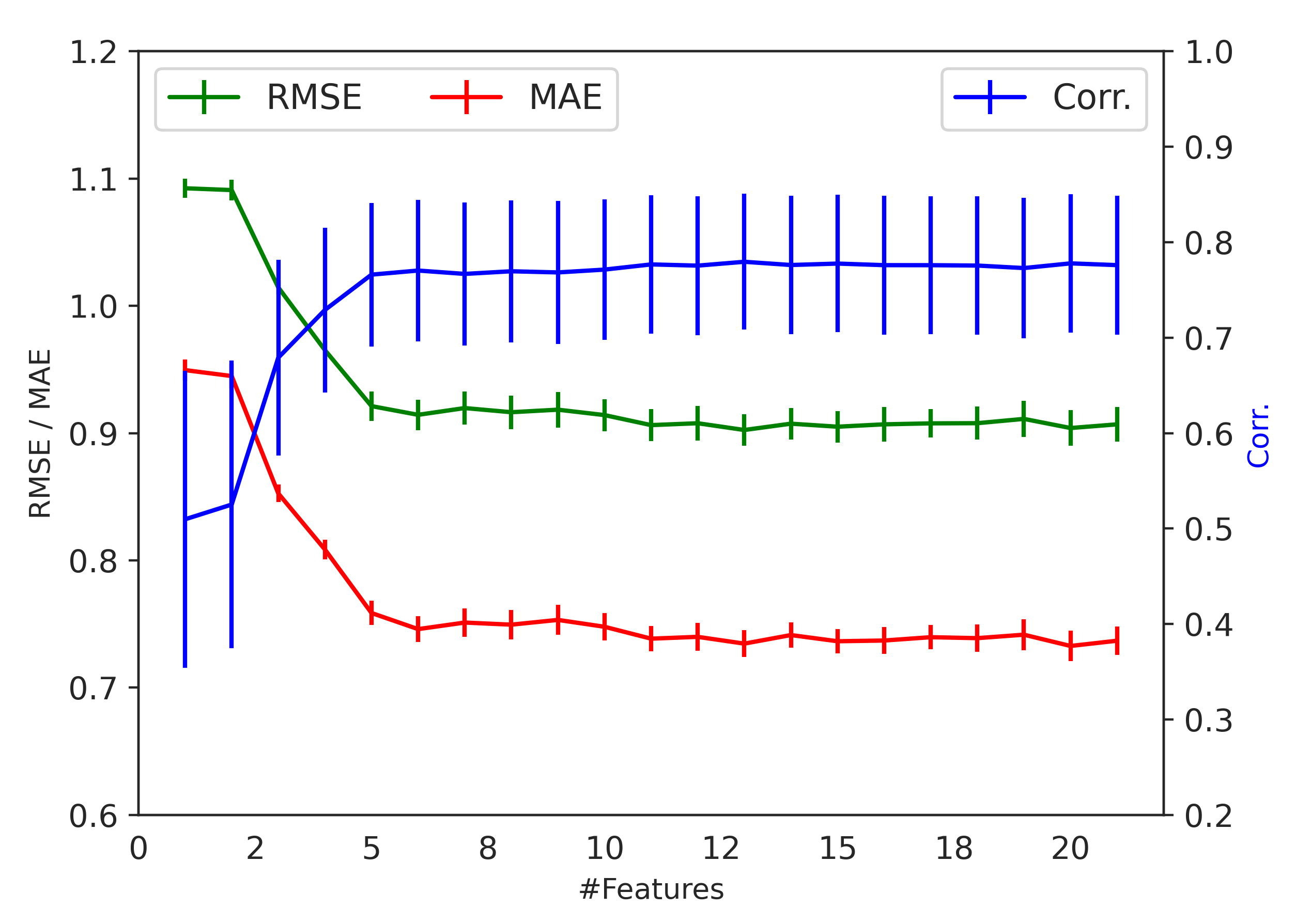}\hfill
\caption{LightGBM model performance over the iteration of adding important featured at the time.}
\label{fig:rmse_mae}
\end{figure}

The LightGBM model's performance in predicting SA is presented in Table \ref{tab:mPerformance}. To analyze how individual features contribute to predictions, SHAP values were calculated and visualized in a SHAP summary plot (Fig. \ref{fig:SHAP}). This reflects both the importance and the effects of each feature. The y-axis positioning is determined by the feature importance, ranging from the most to the least significant.  The x-axis is determined by the SHAP value where very point on the plot represents a feature's SHAP value for a given instance. The color scale, ranging from blue (low) to red (high), indicates the magnitude and direction of a feature's impact on the predicted SA score, and the overlapping points, jittered in y-axis direction, describe the distribution of the SHAP values per feature.
To further optimize the model's performance, the model's behavior was tested by adding the features incrementally according to their SHAP importance rankings. The impact of each addition on the model's accuracy is illustrated in Fig. \ref{fig:rmse_mae}, which shows the the variations in performance metrics (i.e., RMSE, MAE and Corr) with the inclusion of more predictors. It was observed that the LightGBM model resulted in improved performance with a subset of top 12 features rather than the entire feature set. Following this insight, the model was retrained with these 12 features where the updated model's had the best performance (see Table \ref{tab:mPerformance} under the 'Selected features' row).

According to SHAP and model performance, the following features had the most impact on the model and are listed in descending order:  age, avKnowledge, mean\_gsr, mean\_HR, mean\_HRV, number\_of\_fixations\_game, number\_of\_fixations\_center, mean\_dispersion\_game, gender, mean\_duration\_game, mean\_duration\_center, mean\_dispersion\_center. However, the summary plot shows only the global view of how feature values influenced predictions.


\begin{table}[H]
\caption{Performance of LightGBM regressor.}
\label{tab:mPerformance}
\centering
\begin{tabularx}{0.7\linewidth}{llll}
\hline
Sample size & RMSE & MAE & Corr \\ \hline
All features  & 0.90 & 0.72 & 0.77  \\
Selected Features  & 0.89 & 0.71 & 0.78 \\ \hline
\end{tabularx}%
\end{table}

\subsection{Feature Effect in Predicting SA}
\begin{figure}[h!]
\centering
\includegraphics[width=\linewidth]{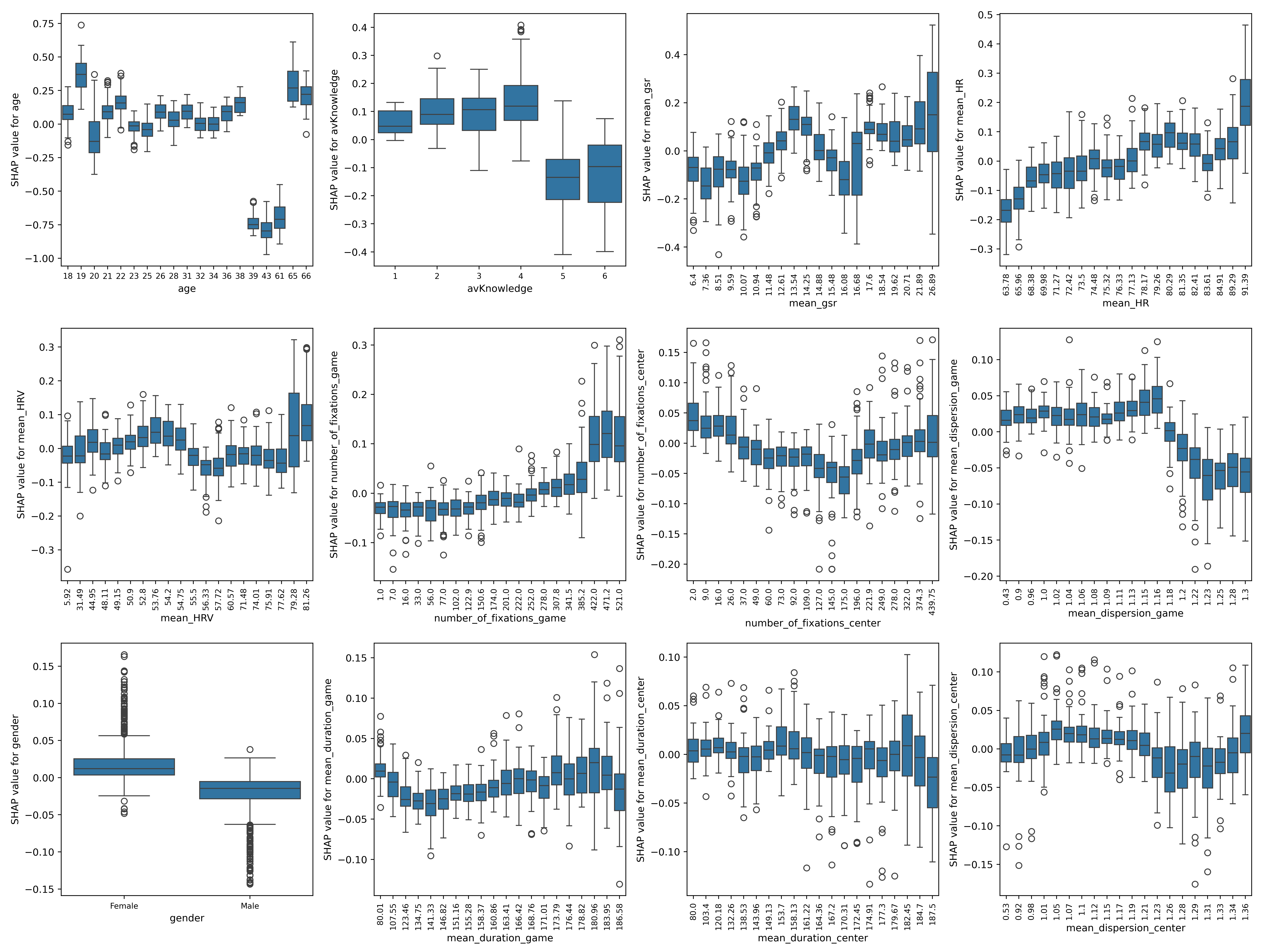}\hfill
\caption{The effect of important features on predicted SA value. The x-axis represents the value of the feature, and the y-axis represents the SHAP value associated with that feature. Positive SHAP values indicate that the feature pushes the prediction higher, while negative values indicate the opposite.}
\label{fig:boxplots}
\end{figure}

To explore the impact of an individual feature on the SA predictions made by the model, for the top 12 features, the SHAP value was charted against feature's actual values. This relationship is shown in Fig. \ref{fig:boxplots}, where the feature values were segmented into several groups according to their distribution (the physiological features were limited up to 20 segments). This segmentation allowed the examination of the feature's effect on SA. Spearman’s rank correlation coefficient for each feature was calculated to understand the strength of the relationship.

Age: The most important feature was age in Fig. \ref{fig:SHAP} with a significant negative correlation ($\rho= -0.294, p < 0.001$).  A complex relationship with SA was observed. Younger adults (18-38) showed minimal influence of age on SA.  A strong negative effect was found for middle-aged adults (39-62) (i.e., those in this age range had significant lower predicted SA), while older adults (62+) displayed a positive effect. However, the dataset's skewed age distribution warrants caution in interpretation.

AV Knowledge: Participants with higher AV knowledge tended to have lower SA predictions ($\rho= -0.665, p < 0.001$).

Physiological Signals: GSR: The mean\_gsr was positively linked to SA ($\rho= 0.463, p < 0.001$), with higher values suggesting an increase in SA, though the relationship was not strictly linear.
Heart Rate: mean\_HR showed a positive relationship with SA ($\rho= 0.670, p < 0.001$), indicating that elevated HR was associated with higher levels of SA. 
HRV: Although the correlation was significant ($\rho= 0.128, p < 0.001$), the effect size suggests a very weak association with no overall trend.

Eye-Tracking Metrics: Fixations (Game): A significant positive correlation was found between the number\_of\_fixations\_game and SA ($\rho= 0.724, p < 0.001$), suggesting that an increased frequency of fixations correlates with enhanced SA.
Fixations (Center): There was a notable pattern characterized by a a decrease in SA with a lower frequency of fixations at the center($\rho= -0.302, p < 0.001$).
Dispersion (Game): The mean\_dispersion\_game seemed to be negatively correlated with SA  ($\rho= -0.233, p < 0.001$), indicating that greater distances between fixation points  on game surface were associated with lower levels of SA.
Dispersion (Center): The mean\_dispersion\_center ($\rho= 0.310, p < 0.001$) seemed to be positively correlated with higher dispersion on center leading to increase of SA.
Fixation durations: Mean duration of fixations on both the game and center surfaces showed weaker correlations, with less clear trends.

Gender: The data showed that female participants tend to have higher SA than males ($\rho= -0.670, p < 0.001$).

\subsection{SA, Trust and Perceived Risk Across Conditions}
A mixed two-way analysis of variance (ANOVA) was used to analyze the effects of risk perception and automation error on participants SA, trust, risk and physiological responses. 
The ANOVA showed a significant main effect of automation error ($F(1,56) = 5.313, p = 0.025, \eta_p^2 = 0.087$) and marginal main effect of risk condition ($F(1,56) = 3.438, p = 0.069, \eta_p^2 = 0.058$) on SA. Within the high-risk group, participants reported a significantly higher level of SA ($p = 0.018$) during the drive with automation error compared to low-risk group.  In terms of trust, a significant main effect of automation error was found ($F(1,62) = 13.700, p < 0.001, \eta_p^2 = 0.181$). For risk perception, although no significant differences were found between the two conditions, it successfully elicit different levels of self-reported SA.

\subsection{Objective Responses Across Conditions}
The effect of risk and automation error was also investigated for physiological features using a two-way mixed ANOVA. 
There was a significant interaction effect on mean\_HR ($F=7.348, p=0.012$), with a large effect size ($\eta_p^2=0.242$).
For mean\_duration\_center, there were significant main effects of both risk ($F=5.34, p=0.03, \eta_p^2=0.188$) and automation error ($F=9.077, p=0.006, \eta_p^2=0.283$), as well as a significant interaction effect ($F=5.089, p=0.034, \eta_p^2=0.181$). The pairwise t-test showed that drive with automation errors resulted in longer fixation duration in the center compared to drive where no error was experienced. Additionally, the high risk condition led to longer center fixation duration than the low risk condition. 
Mean\_dispersion\_center showed significant main effects of risk  ($F=5.988, p=0.022, \eta_p^2=0.207$) and automation error ($F=4.657, p=0.042, \eta_p^2=0.168$), where automation error increased dispersion in the center area, similarly, the high risk environment increased dispersion compared to low risk.
The automation error had significant main effect on number\_of\_fixations\_center ($F=10.786, p=0.003, \eta_p^2=0.319$) where fixation number was higher when AV had errors.
For the number\_of\_fixations\_game automation error showed significant main effect as well($F=8.423, p=0.008, \eta_p^2=0.268$) where participants had more fixations on game display during the drives without automation error.
Risk significantly impacted on mean\_dispersion\_left ($F=5.516, p=0.028, \eta_p^2=0.193$) and mean\_dispersion\_right ($F=4.41, p=0.047, \eta_p^2=0.161$), with high risk leading to greater dispersion compared to low risk. Finally, automation error had significant main effects on mean\_duration\_odometer ($F=4.698, p=0.041, \eta_p^2=0.17$) and mean\_dispersion\_odometer ($F=6.817, p=0.016, \eta_p^2=0.229 $), with higher duration and dispersion when automation had errors.

\begin{figure}[h!]
\centering
\includegraphics[width=\linewidth]{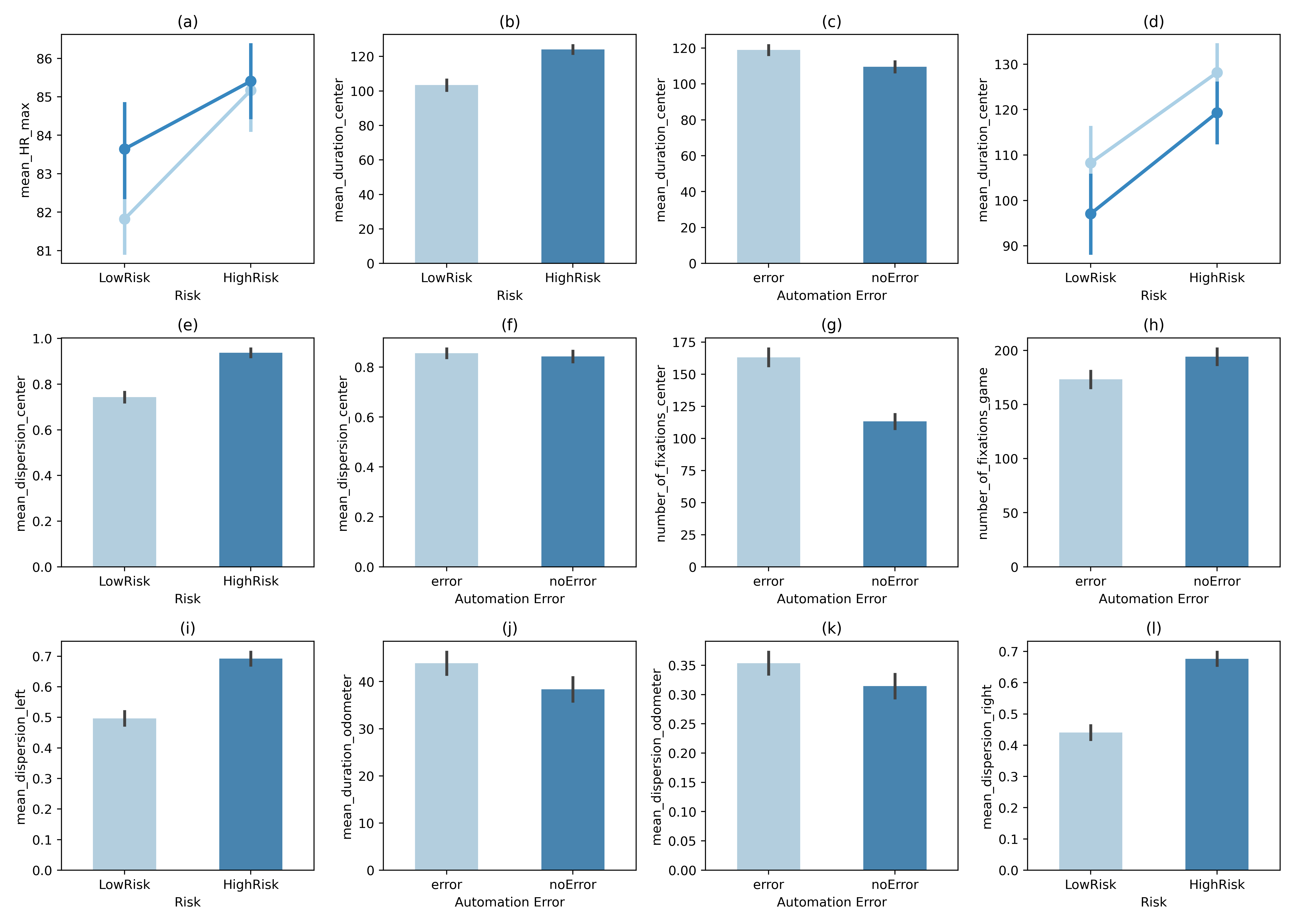}\hfill
\caption{Main effects on features across tested conditions.}
\label{fig:sig_plots}
\end{figure}

\section{Discussion}


\subsection{Predicting SA with machine learning model}
In our investigation, the LightGBM machine learning model has demonstrated promising capabilities in estimating SA from a diverse array of signals. By incorporating a multimodal dataset—including physiological signals, eye tracking metrics, and demographic information—the model was able to predict SA levels with RMSE of 0.90, MAE of 0.72, and a correlation coefficient of 0.77 with self-reported SA measures.

To contextualize these results, we employed the SHAP (SHapley Additive exPlanations) framework to interpret the predictive model. The SHAP summary plot not only illuminated the overarching importance of each feature but also revealed the nature of their effects on SA prediction. It is noteworthy that we identified certain features whose contribution appeared marginal, potentially serving as noise that detracted from the model accuracy \citep{ayoub2023real}. When considering a refined feature set with the top 12 variables as indicated by their SHAP values, the model's performance was marginally enhanced, showing an RMSE of 0.89, an improved MAE of 0.71, and a boosted correlation of 0.78.

Among the principal features influencing the model's predictions were demographic aspects such as age, gender and experience with automated vehicles. This seemed to be consistent with previous findings that demographic variables could reflect diverse cognitive abilities and and confidence levels in tasks, thereby impacting SA in an automated context \citep{kintz2023estimation, li2018investigation}. 

Physiological signals, including GSR, HR, and HRV, were also important. Mean\_gsr was positively correlated to SA, possibly because a higher level of GSR was related to a high level of arousal and alertness \citep{zhou2011affect, zhou2014emotion}, which further led to a higher level of arousal. Mean\_HR was also positively correlated with SA, which might be a pertinent indicator of participants' stress levels reacting to potential risks and errors of the AV, which was integral for the participants to pay more attention to the driving situations \citep{perello2022physiological}. Previous studies also found that a higher value of HRV was associated with better attentional maintenance and flexibility, which might contributed to its positive correlation with SA \citep{siennicka2019resting}. 


We found a positive correlation between the number of fixations on the game display and SA and a negative correlation between the number of fixations on the center area (presumably the road ahead) and SA. Such findings are somewhat counterintuitive. Previous studies showed that increased visual attention towards the NDRTs (i.e., game) suggested a lack of focus on the driving environment, potentially leading to lower SA and typically more frequent fixations on the road were associated with better SA \citep{zhou2021using, liang2021look}. However, such findings could be influenced by other factors, such as the duration or timing of fixations, which may provide more insights into the driver's attentional allocation and comprehension of the driving situation.

The negative correlation between the dispersion of fixations on the game and SA was consistent with the idea that excessive visual engagement with NDRT degraded SA \citep{zhou2021using, du2020psychophysiological}. Greater dispersion on the game surface may indicate increased distraction and reduced focus on the driving environment, leading to lower SA.
The positive correlation between dispersion of fixations on the center area and higher SA aligned with previous findings \citep{du2020psychophysiological, liang2021using}. A wider distribution of fixations on the road ahead and surrounding areas can facilitate better perception and comprehension of the driving situation, thereby enhancing SA. 

Overall, the integration of machine learning techniques with multimodal datasets presents a powerful approach to uncover the nuanced factors that influence SA in the realm of automated driving. Through continual refinement of feature sets and comparison with extant research, we might not only be able to advance the predictability of SA levels, but also understand the underlying cognitive and behavioral processes involved.

\subsection{Effects of risks and automation errors on SA}

In this study, we manipulated participants' risk perceptions through exposure to risk-related content and varied the performance of the automated driving system and the simulated driving environment. The results revealed that when participants were exposed to high-risk content, their SA levels were higher. These findings are supported by previous studies who demonstrated that (\citep{li2019no, yahoodik2021effectiveness}) perceived risks and trust in an AV were affected by introductory information and received training related to system reliability. 

Notably, SA was significantly sensitive to automation performance and was higher when the system experienced failures compared to error-free drives. This observation is consistent with studies indicating that automation errors and system limitations can trigger compensatory behaviors and heightened alertness in drivers, leading to improved SA during critical situations \citep{korber2018introduction, schwarz2019} 

The eye tracking responses supported this observation, as participants spent a significant amount of time gazing at the road center when driving in risk conditions (see Fig. \ref{fig:sig_plots} b) and experienced automation errors (see Fig. \ref{fig:sig_plots} c). Moreover, the high risk condition resulted in notably increased dispersion on road center (see Fig. \ref{fig:sig_plots} e) specifically in erroneous situations (see Fig. \ref{fig:sig_plots} f), as well as left and right sides (see Fig. \ref{fig:sig_plots} i and l). This broader visual scanning pattern suggests that individuals may explore the traffic environment more extensively and scan across a wider range of locations when faced with higher perceived risks or system failures, potentially in an attempt to gather more information and enhance their understanding of the situation \citep{thill2014apparent, de2019situation}.

Furthermore, the data from the eye tracker revealed that participants checked the traffic around them more frequently in presence of automation error (see Fig. \ref{fig:sig_plots} g) and were less focused on the NDRT screen (see Fig. \ref{fig:sig_plots} h). This shift in attentional allocation from the NDRT to the road scene during automation errors is consistent with previous findings that drivers tend to prioritize monitoring the driving environment over secondary tasks when faced with critical situations or system limitations \citep{ smith2023physiological, naujoks2014effect}.
 
Finally, we observed that the participants demonstrated longer and more dispersed gazing on the odometer when they experienced failures (see Fig. \ref{fig:sig_plots} j and k), which was noted when participants noticed the hazard ahead before they received a response from the AV. This behavior might indicate that they were trying to understand if the automation was going to react to the potential hazard or checking if the automation was turned off after they resumed control \citep{zeeb2015determines, niu2022driver}.

Overall, these findings contribute to our understanding of how risk perception and automation reliability influence drivers' visual attention allocation, situation scanning strategies, and cognitive processes related to SA in conditionally automated driving scenarios. The results align with and extend the existing literature on the effects of perceived risk, system limitations, and critical events on drivers' vigilance, compensatory behaviors, and SA in human-automation interaction contexts.

\subsection{Implications}
We developed a context-aware model using machine learning to predict SA during conditionally AVs, which could have significant implications for enhancing safety and user experience. 
First, by combining diverse data sources like demographics, physiological signals, and eye movements, the model can capture a more comprehensive understanding of the driver's state and attentional focus. This holistic approach could lead to more accurate predictions of SA levels during automated driving scenarios, which could provide deeper insights into the driver's cognitive processes and readiness to take over control when needed \citep{du2019examining}. Moreover, incorporating demographic factors like age, gender, and driving experience into the model allows for personalized driver monitoring and tailored interventions \citep{avetisyan2023building}.

By accurately predicting SA levels, the model can inform the design of adaptive in-vehicle systems and human-machine interfaces (HMIs) that provide context-aware support and warnings to drivers \citep{pakdamanian2022enjoy}. This could lead to safer and smoother transitions between automated and manual driving modes.
The model's predictions can also guide the development of personalized training programs or adaptive automation strategies, helping drivers maintain an appropriate level of SA and readiness during automated driving \citep{du2019examining}.

Overall, developing a context-aware ML model that leverages multimodal data sources has the potential to significantly enhance SA prediction and support safer and more user-friendly conditionally automated driving experiences. However, technical challenges, data availability, and ethical considerations must be carefully addressed to realize the full benefits of this approach.

\subsection{Limitations and Future Work} 
This study has several limitations that should be acknowledged. First, the experimental setup used a low-fidelity desktop driving simulator, which may not fully replicate the realistic dynamics and risk perceptions of an on-road driving environment. Future studies should aim to conduct experiments in higher-fidelity simulated or real-world settings to enhance the ecological validity of the findings.

Second, the takeover scenarios tested in this study covered a limited range of risk levels. To gain a more comprehensive understanding of SA level, it is essential to explore a broader spectrum of risky situations, including more critical and time-sensitive takeover events.

Third, the study relied on a single self-reported item to measure SA, which may not fully capture the multidimensional nature of this construct. Future research should explore alternative methods for assessing situation awareness, such as objective performance measures or more comprehensive self-report instruments, to establish a more reliable ground truth.

Finally, integrating diverse data sources and developing robust ML models for SA prediction can be technically challenging, requiring advanced data fusion techniques and large, high-quality datasets for training and validation. The model's performance and reliability in real-world driving scenarios need to be thoroughly evaluated, as factors like environmental conditions and unexpected events can influence SA and driver behavior.

\section{Conclusions} 
This paper presents research on developing a predictive model for assessing SA in conditionally automated driving scenarios. An experiment with 67 participants using a driving simulator was conducted. Participants experienced automated driving with TOR events, including some with automation errors (i.e., false alarms and misses). Their physiological responses (GSR, HR, eye tracking) and self-reported SA were recorded.
The LightGBM machine learning model was used to predict SA levels from the physiological and demographic data. The model achieved reasonable performance ($RMSE = 0.89, MAE = 0.71, Corr = 0.78$) using a subset of the top 12 most important features resulted from SHAP explainer. The key findings were: 1) age, AV knowledge, GSR, HR, and eye behavior on the center and NDRT screens were the most influential predictors of SA, 2) higher risk perception led to larger fixations durations and dispersions on center screen,  3) automation errors increased the dispersions and fixations on  center and NDRT screens, and 4) SA was higher during automation error conditions for the high risk group compared to low risk.

These findings contribute to our understanding of the factors influencing SA in conditionally automated driving scenarios, particularly considering the impact of risk perception and automation errors. The developed predictive model demonstrates the potential for using physiological, behavioral and demographic measures to monitor and assess drivers' SA in real-time, enabling intelligent vehicle systems to provide timely interventions or explanations to enhance SA and promote safer human-AV interactions.


\section{Acknowledgement}
This research was supported by National Science Foundation.

\bibliography{main}
\newpage
\section{Biographies}
\includegraphics[width=1in,height=1.55in,clip,keepaspectratio]{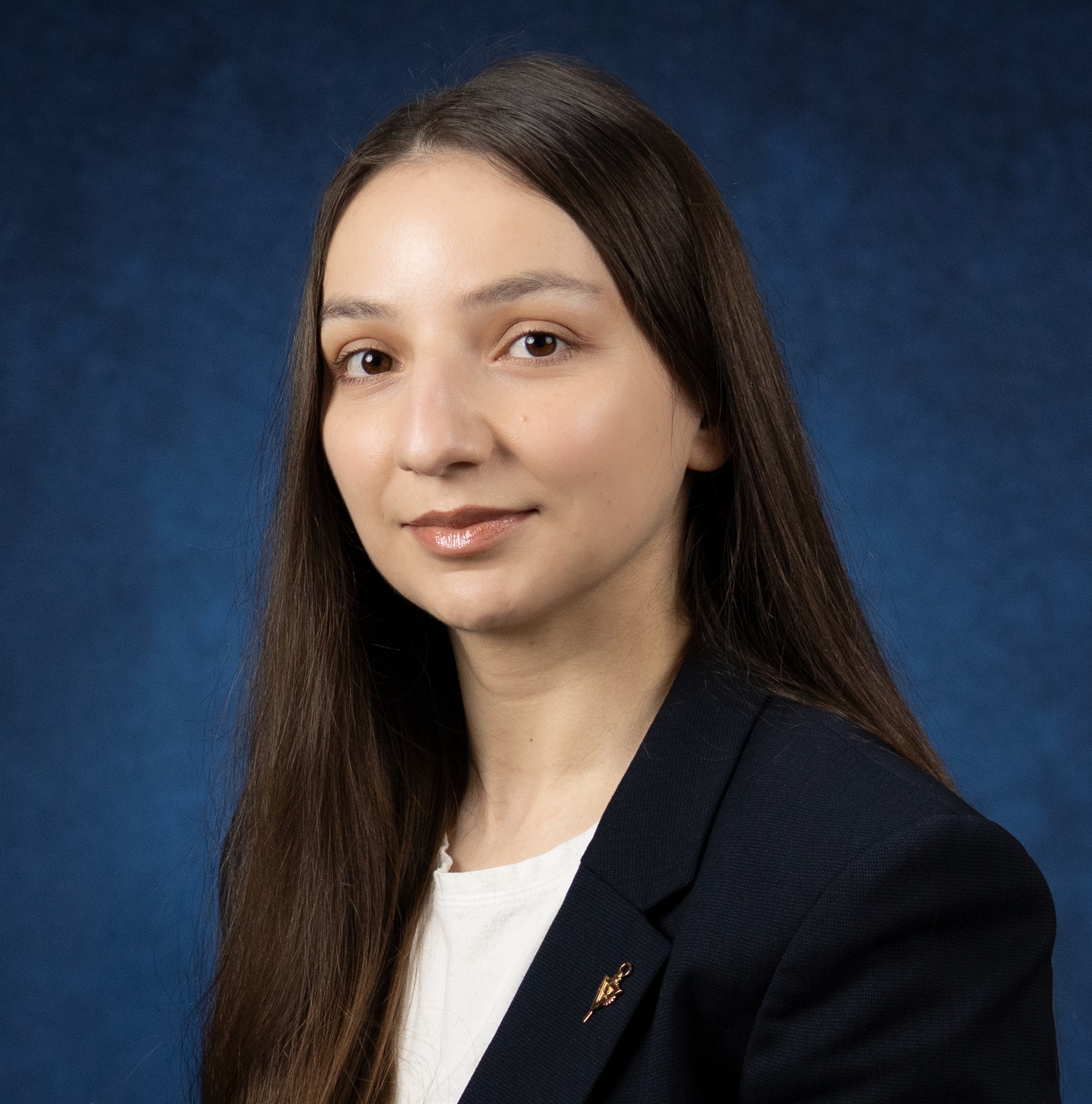}
\textbf{Lilit Avetisyan} received her B.E. degree in 2017 and MS degree in 2019 in Information Security from the National Polytechnic University of Armenia. She is currently pursuing her Ph.D. degree in Industrial and Systems Engineering at the University of Michigan, Dearborn. Her main research interests include human-computer interaction, explainable artificial intelligence and situation awareness.\\

\includegraphics[width=1in,height=1.55in,clip,keepaspectratio]{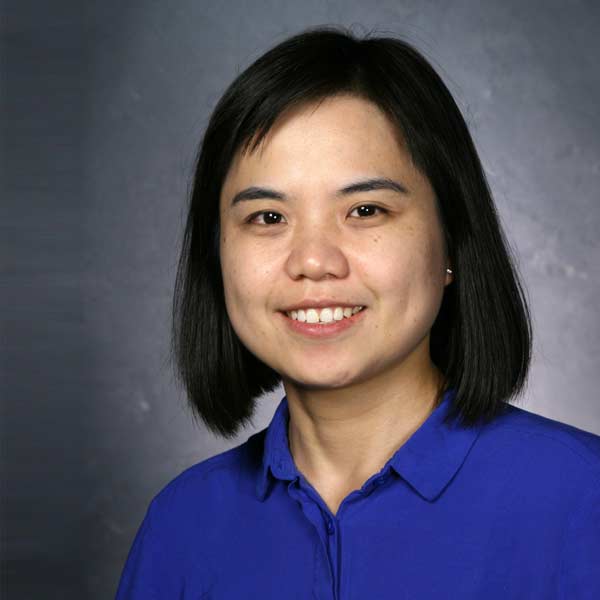}
\textbf{X. Jessie Yang} is an Assistant Professor in the Department of Industrial and Operations Engineering, University of Michigan, Ann Arbor. She earned a PhD in Mechanical and Aerospace Engineering (Human Factors) from Nanyang Technological University, Singapore. Dr. Yang’s research include human-autonomy interaction, human factors in high-risk industries and user experience design.\\

\includegraphics[width=1in,height=1.55in,clip,keepaspectratio]{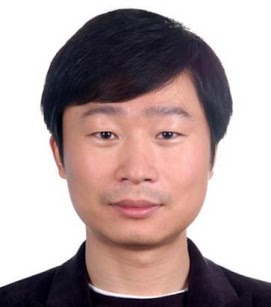}
\textbf{Feng Zhou} received the Ph.D. degree in Human Factors Engineering from Nanyang Technological University, Singapore, in 2011 and Ph.D. degree in Mechanical Engineering from Gatech Tech in 2014. He was a Research Scientist at MediaScience, Austin TX, from 2015 to 2017. He is currently an Assistant Professor with the Department of Industrial and Manufacturing Systems Engineering, University of Michigan, Dearborn. His main research interests include human factors, human-computer interaction, engineering design, and human-centered design.\\

\end{document}